\documentclass[format=acmtog, screen=true, authorversion=true]{acmart}

\setcopyright{acmlicensed}
\copyrightyear{2020}

\acmJournal{TOG}
\acmYear{2020}
\acmVolume{39}
\acmNumber{4}
\acmArticle{88}
\acmMonth{7}
\acmDOI{10.1145/3386569.3392444}

\acmSubmissionID{papers\_388}

\usepackage[noannotation]{latexie}

\newcommand{\figteaser}{
  \begin{teaserfigure}
    \centering
    \includegraphics[width=\textwidth]{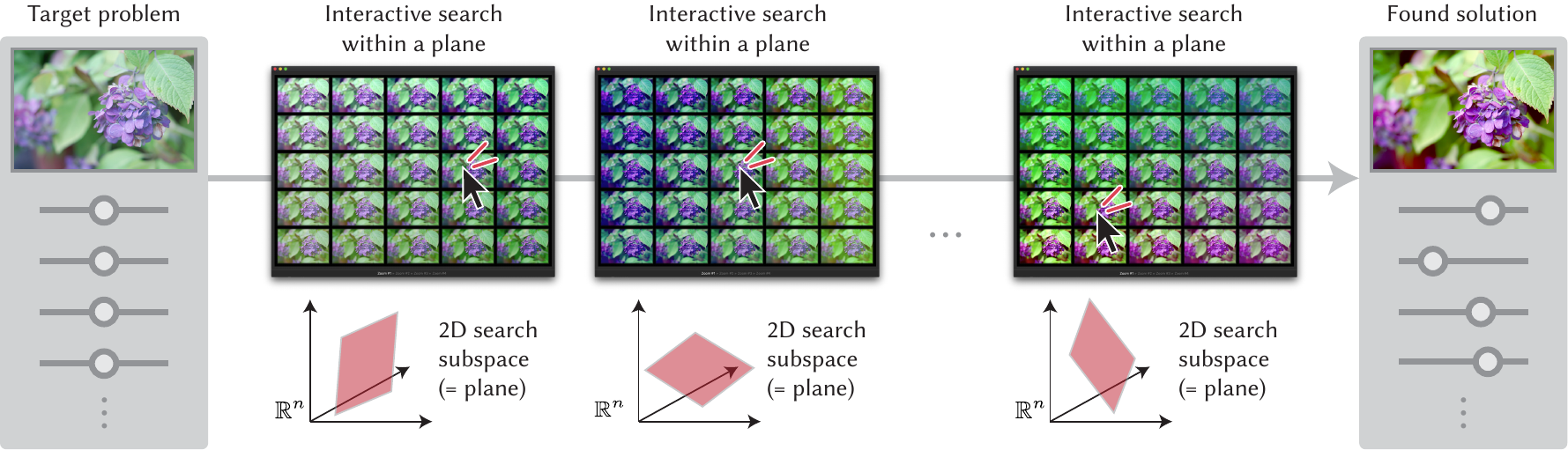}
    \caption{
      \textbf{\emph{Sequential Gallery}} is an interactive framework for exploring an $n$-dimensional design space formed by a set of $n$ sliders and then finding an appropriate parameter set from that space.
      This framework lets the user sequentially select the most preferable option from the options displayed in a grid interface.
      To enable this framework, we propose a new \emph{Bayesian optimization} method called \emph{sequential plane search}, which decomposes the original high-dimensional search problem into a sequence of two-dimensional search (\ie, plane-search) subtasks.
    }
    \vspace{1mm}
    \label{fig:teaser}
    \Description{}
  \end{teaserfigure}
}

\newcommand{\figmapping}{
  \begin{figure}
    \centering
    \includegraphics[width=\columnwidth]{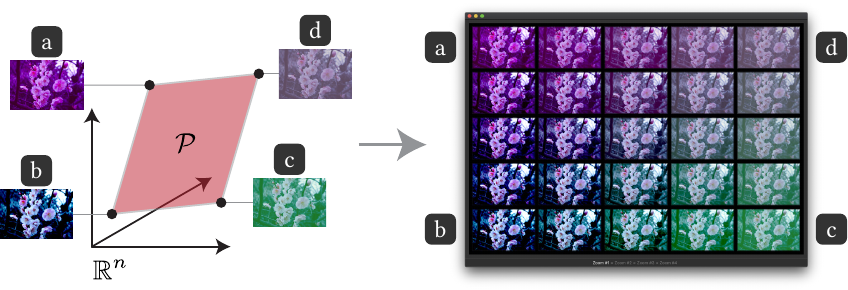}
    \caption{
      \textbf{Illustration of the mapping} from a search plane $\calP$ in the target $n$-dimensional design space to the zoomable grid interface.
    }
    \label{fig:mapping}
    \Description{}
  \end{figure}
}

\newcommand{\figzoom}{
  \begin{figure*}
    \centering
    \includegraphics[width=\textwidth]{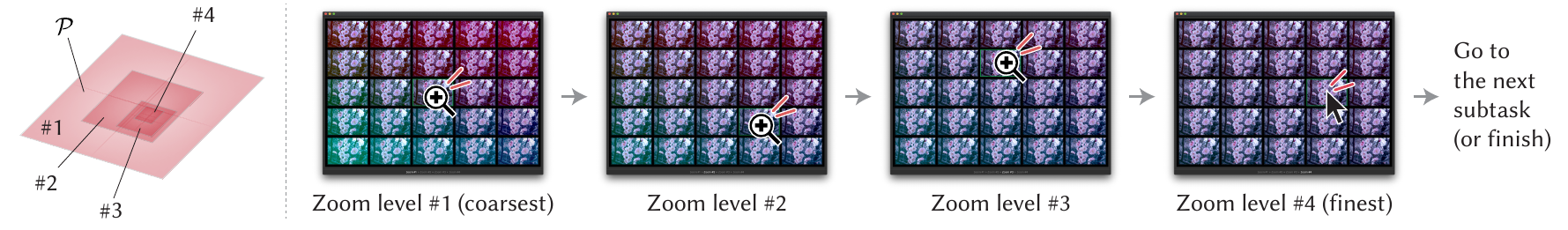}
    \caption{
      \textbf{Zooming procedure in the zoomable grid interface.}
      The user clicks the best option displayed in the grid, and then the interface goes to the next finer zoom level. After a certain number of clicks (four in this example), this plane-search subtask ends. If the user wants to continue exploration, our method constructs a new search plane and then asks the user to start another zooming procedure.
    }
    \label{fig:zoom}
    \Description{}
  \end{figure*}
}

\newcommand{\figplanarity}{
  \begin{figure}
    \includegraphics[width=\columnwidth]{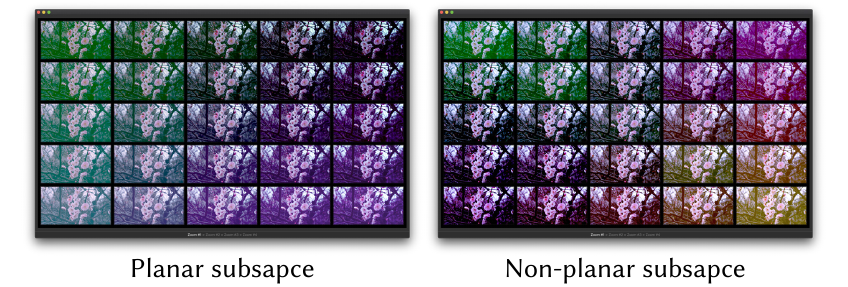}
    \caption{
      \textbf{Importance of planarity of the search subspace.}
      When the search subspace is planar (Left), the grid exhibits linear parameter changes across all directions. In many cases, this makes it easier for users to grasp the current plane than when the search subspace is not planar (Right).
    }
    \label{fig:planarity}
    \Description{}
  \end{figure}
}

\newcommand{\fignotation}{
  \begin{figure}
    \centering
    \includegraphics[width=\columnwidth]{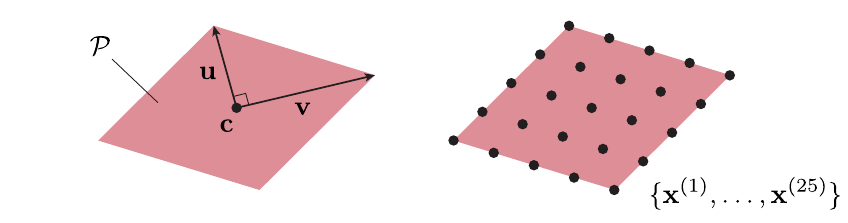}
    \caption{
      (Left)
      \textbf{Parameterization} of a plane $\calP$.
      (Right)
      \textbf{Sampling points} used in our implementation for approximating the surface integral (\autoref{eq:acquisition:integral}).
    }
    \label{fig:notation}
    \Description{}
  \end{figure}
}

\newcommand{\figphotofull}{
  \begin{figure*}
    \centering
    \includegraphics[width=\textwidth]{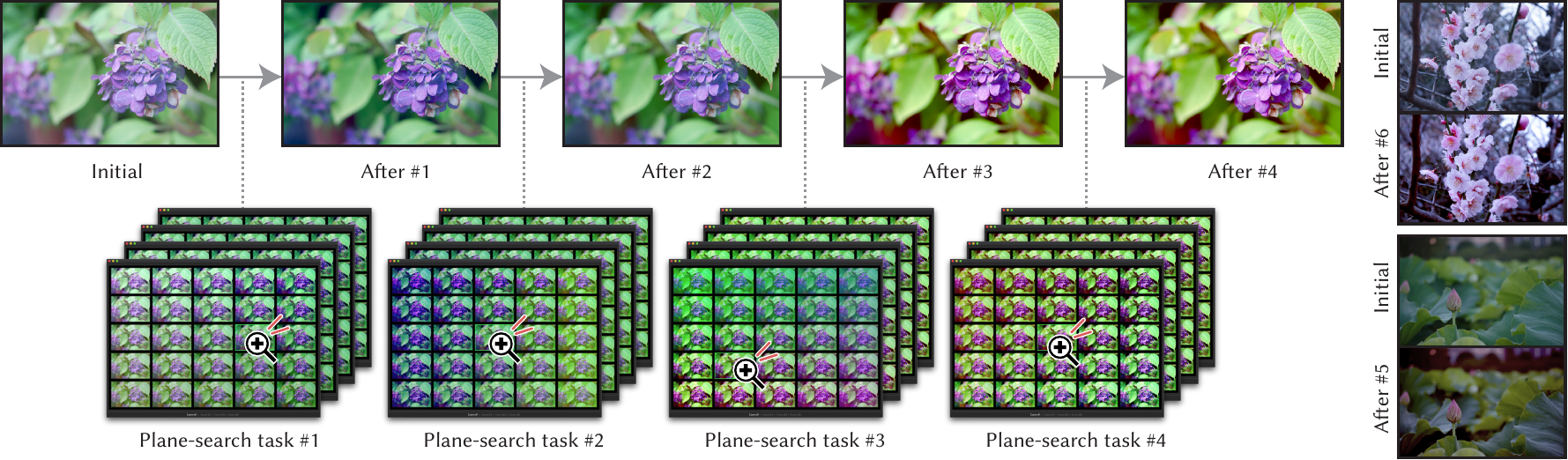}
    \caption{
      \textbf{Photo color enhancement with Sequential Gallery}, in which 12 design parameters were adjusted.
      (Left) An entire optimization sequence, in which the user could obtain a satisfactory result after four iterations.
      (Right) Additional results; refer to the supplemental video figure for details.
    }
    \label{fig:photo-full}
    \Description{}
  \end{figure*}
}

\newcommand{\figsmplspace}{
  \begin{figure}
    \includegraphics[width=\columnwidth]{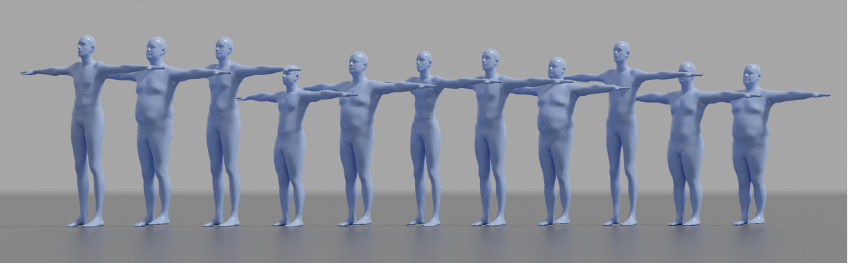}
    \caption{
      \textbf{Random human body shapes} generated from the SMPL model \protect\cite{LoperSA15}.
      We used the top-10 dimensions as the design space.
    }
    \label{fig:smpl-space}
    \Description{}
  \end{figure}
}

\newcommand{\figbodyresults}{
  \begin{figure}
    \includegraphics[width=\columnwidth]{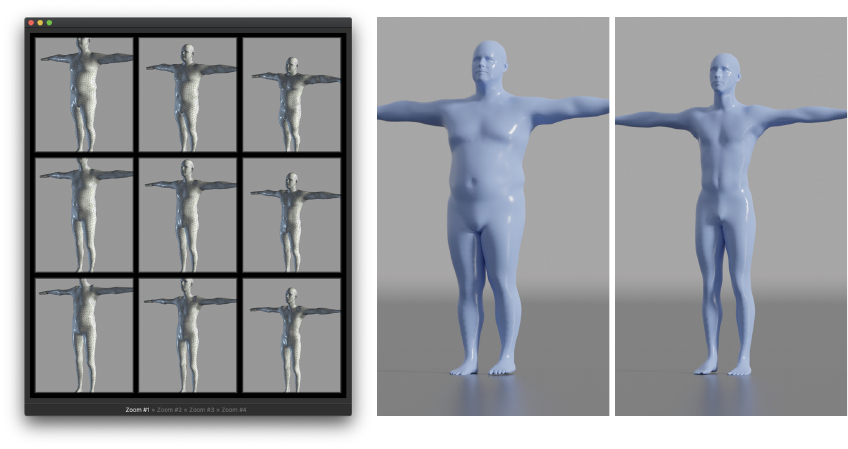}
    \caption{
      (Left)
      \textbf{Appearance of the zoomable grid interface in the human body shape design scenario,}
      where we set the grid resolution as 3 by 3.
      (Center)
      \textbf{Body shaping from a description of a character in a novel},
      \emph{The Maltese Falcon} \protect\cite{Falcon}.
      (Right)
      \textbf{Body shaping of a famous fictional character},
      \emph{Spider-Man} \protect\cite{SpiderMan}.
    }
    \label{fig:body-results}
    \Description{}
  \end{figure}
}

\newcommand{\figsynthetic}{
  \begin{figure*}
    \includegraphics[width=\textwidth]{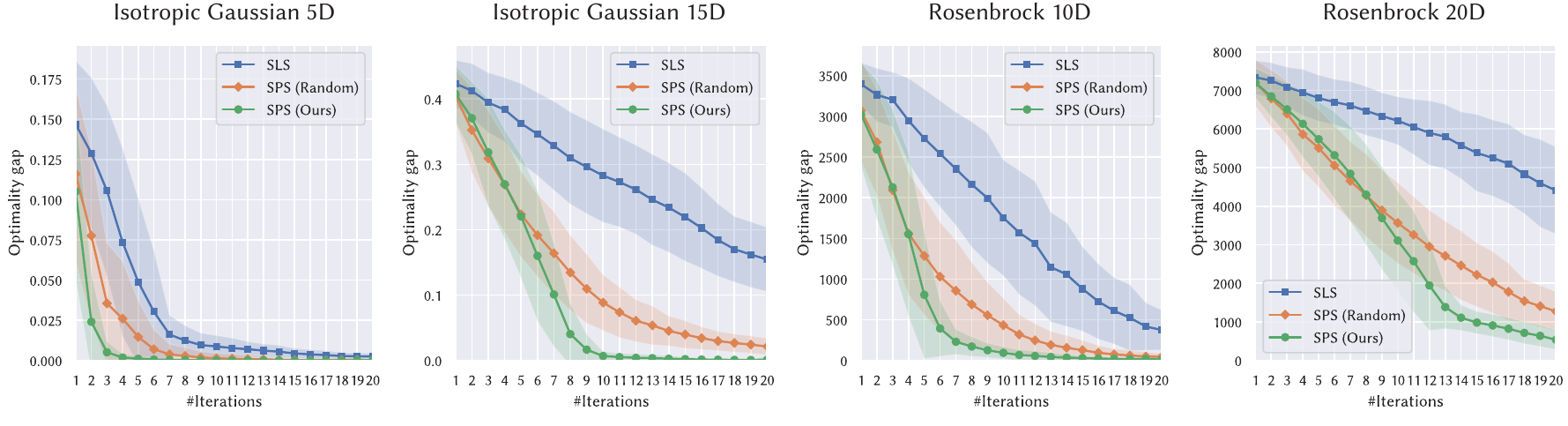}
    \caption{
      \textbf{Results of the experiment with synthetic functions.}
      We compare the sequential line search (SLS) \protect\cite{Koyama:SIGGRAPH:17}, the sequential plane search (SPS) using random plane construction, and our SPS using Bayesian optimization-based plane construction. We run 50 trials for each condition. Each plot shows the mean value with the colored regiong showing the standard deviation. Vertical axes represent optimality gaps (lower is better).
    }
    \label{fig:synthetic}
    \Description{}
  \end{figure*}
}

\begin{document}

\title{Sequential Gallery for Interactive Visual Design Optimization}

\author{Yuki Koyama}
\email{koyama.y@aist.go.jp}
% \orcid{}
\affiliation{%
  \institution{National Institute of Advanced Industrial Science and Technology (AIST)}
  \streetaddress{1-1-1 Umezono}
  \city{Tsukuba}
  \state{Ibaraki}
  \country{Japan}
  \postcode{305-8568}
}
\author{Issei Sato}
\email{sato@k.u-tokyo.ac.jp}
% \orcid{}
\affiliation{%
  \institution{The University of Tokyo}
  \streetaddress{Hongo 7-3-1}
  \city{Bunkyo}
  \state{Tokyo}
  \country{Japan}
  \postcode{113-8654}
}
\author{Masataka Goto}
\email{m.goto@aist.go.jp}
% \orcid{}
\affiliation{%
  \institution{National Institute of Advanced Industrial Science and Technology (AIST)}
  \streetaddress{1-1-1 Umezono}
  \city{Tsukuba}
  \state{Ibaraki}
  \country{Japan}
  \postcode{305-8568}
}

\begin{abstract}
  % !TEX root = ../paper.tex

Visual design tasks often involve tuning many design parameters.
For example, color grading of a photograph involves many parameters, some of which non-expert users might be unfamiliar with.
We propose a novel user-in-the-loop optimization method that allows users to efficiently find an appropriate parameter set by exploring such a high-dimensional design space through much easier two-dimensional search subtasks.
This method, called \emph{sequential plane search}, is based on \emph{Bayesian optimization} to keep necessary queries to users as few as possible.
To help users respond to \emph{plane-search} queries, we also propose using a gallery-based interface that provides options in the two-dimensional subspace arranged in an adaptive grid view.
We call this interactive framework \emph{Sequential Gallery} since users sequentially select the best option from the options provided by the interface.
Our experiment with synthetic functions shows that our sequential plane search can find satisfactory solutions in fewer iterations than baselines.
We also conducted a preliminary user study, results of which suggest that novices can effectively complete search tasks with Sequential Gallery in a photo-enhancement scenario.

\end{abstract}




\figteaser

\maketitle

% !TEX root = ../paper.tex

\section{Introduction}

Visual design tasks often involve many parameters that should be carefully adjusted via sliders.
The purpose of tweaking these parameters is, for example, to reproduce the desired design in mind or to make the design as aesthetically pleasing as possible.
This process is, however, often difficult because the parameters may affect the design in combination and the space of possible parameter configurations is very broad due to the high dimensionality.
Moreover, evaluating a certain parameter configuration is also difficult without actually manipulating the slider values and seeing the corresponding visual representation, which thus requires many trials and errors.
All this is especially true when users are unfamiliar with the design parameters.
For example, photo retouch software has many sliders for color enhancement, including advanced ones such as ``shadows (red)'' and ``highlights (red)'' \cite{AdobePhotoshopColorBalance,InstagramEffects}, which both affect shades of red but in different ways and can produce various effects in combination with other parameters.
This complexity requires users to try many slider configurations at the beginning to understand what effects are possible and tweak the slider values little by little alternately in the last fine-tuning step.
Similar parametric design scenarios appear in many graphics applications, including, but not limited to, material appearance design \cite{McAuleySIGGRAPH12,NganEGSR06}, procedural modeling and animation \cite{Houdini}, rigged character animation \cite{LewisEG14}, personalized fabrication \cite{ShugrinaSIGGRAPH15}, digital composition \cite{Nuke}, and generative design using learned models \cite{YumerUIST15,JinArXiv17}.
Parametric design is also actively used in other domains, such as architecture and product design \cite{Grasshopper}.

We propose a user-in-the-loop optimization method, called \emph{sequential plane search}, that allows users to tactically explore such a high-dimensional design space and efficiently find an appropriate parameter set.
Its novelty is that it decomposes the original search problem into a sequence of much easier two-dimensional search subtasks, called \emph{plane-search} subtasks.
This method is based on \emph{Bayesian optimization} (BO), which is a black-box optimization technique that has recently become popular in the machine learning community \cite{ShahriariProcIEEE16}.
BO automatically balances \emph{exploration} (\ie, encouragement to visit unobserved regions) and \emph{exploitation} (\ie, encouragement to visit high expectation regions) on the basis of Bayesian inference, by which it tries to minimize the number of necessary observations to find a good solution.
By taking advantage of this characteristic in determining plane-search subtasks, our sequential plane search enables users to perform a structured and efficient design exploration guided by a computational strategy.

We also propose an interactive framework, called \emph{Sequential Gallery}, that provides a gallery-based interface to help users effectively perform plane-search subtasks; see \autoref{fig:teaser} for the overview.
The interface is called \emph{zoomable grid} and works as follows.
It displays a finite set of clickable visual options from the search plane in a grid.
At the beginning of a plane-search subtask, it provides a wide variety of options by mapping the entire region of the plane.
Then, it lets users ``zoom'' into a relevant region by selecting the best option among displayed ones.
After a few zooms, this subtask finishes with the best option in the search plane.
Users repeat this coarse-to-fine selection process (\ie, perform the plane-search subtasks sequentially) until they find a satisfactory design.
This gallery-based interface allows users to efficiently grasp possible designs in the subtask without actively manipulating sliders even when they are unfamiliar with the target design space at the beginning of the task.

We demonstrate the generality and applicability of our approach by using two different scenarios:
enhancing colors of photo\-graphs (12 dimensions) and generating human body shapes using a learned generative model (10 dimensions).
To evaluate our sequential plane search, we conducted a simulated experiment using synthetic functions.
This experiment revealed that this method could find good solutions in fewer iterations than baseline methods, including the recently proposed \emph{sequential-line-search} method \cite{Koyama:SIGGRAPH:17}, on top of which our method is built.
We also conducted a preliminary user study with the photo enhancement scenario, which suggested that novices could effectively complete parameter tweaking tasks and produce satisfactory designs with Sequential Gallery.

In summary, our contribution is twofold.
\begin{itemize}
  \item We propose a novel method called \emph{sequential plane search} for user-in-the-loop visual design optimization, which requires fewer iterations to find a satisfactory design parameter set than the previous method \cite{Koyama:SIGGRAPH:17}.
  We evaluate its performance against baselines through simulated experiments.
  \item We use a \emph{zoomable grid} interface in combination with our sequential-plane-search method, which enables users to effectively explore the design space and perform the search.
  We tested this interactive framework, named \emph{Sequential Gallery}, through a small user study.
\end{itemize}

% !TEX root = ../paper.tex

\section{Related Work}

\subsection{Interfaces for Exploration and Parameter Tweaking}

Exploratory design is a design process in which the goal is only loosely specified at the beginning but becomes more and more concrete (or even changes) through exploration \cite{Talton:SA:11}.
To facilitate this process, researchers have investigated gallery-based design interfaces.
The seminal work by Marks \etal\ \shortcite{MarksSIGGRAPH97}, called \emph{Design Gallery}, visually displays representative design options on a two-dimensional screen by low-dimensional embedding.
Followers have demonstrated gallery-based interfaces for effectively exploring complex design spaces of image recoloring \cite{ShapiraEG09} and material reflectance design \cite{NganEGSR06} using domain-specific formulations.
The Brainstorm tool \cite{AdobeBrainstorm} aimed at providing users with inspiration, especially at the beginning of exploration, by showing a gallery of randomly sampled designs.
Lee \etal\ \shortcite{LeeCHI10} reported how a gallery of example designs could help users obtain inspiration to improve design work.
Our grid interface follows these gallery-based approaches since our target scenario is similar to exploratory design (though we do not assume that users' preferences drift over time as described in \autoref{sec:background}).

Another common approach to facilitating parameter tweaking is to augment slider interfaces for more effective direct manipulation.
\emph{Side Views} \cite{TerryUIST02}, a mechanism to augment graphical user interface widgets, can augment sliders with design previews.
\emph{VisOpt Slider} \cite{Koyama:UIST:14,Koyama:CHI:16} visualizes estimated ``goodness'' of slider values using a colormap so that it gently guides users to relevant regions of the target design space during slider manipulation.
Desai \etal\ \shortcite{DesaiCHI19} further extended this interface and used it for robotic motion design.
In contrast to these parameter-wise editing approaches, we take the \emph{what-you-see-is-what-you-get} (WYSIWYG) approach:
users can explore the space by interacting with visual representations without caring about raw parameter values, which eliminates the need for being familiar with (or creating a mental model of) the design parameters.

To help users complete plane-search subtasks, we propose using a zoomable grid interface instead of using two sliders with a preview.
This interface follows the concept of ``zoom-and-pick'' (\eg, \cite{ForlinesUIST05}):
users can select a point from the target space precisely by zooming around the relevant region.

\subsection{Bayesian Optimization and Preference Learning}

Bayesian approaches have been drawing more and more attention in building interactive systems \cite{KristenssonCHIEA19}.
Our sequential-plane-search method is based on one such approach, called Bayesian optimization (BO) \cite{BrochuArXiv10,ShahriariProcIEEE16}, which is a black-box global optimization technique.
BO tries to minimize the number of necessary queries to obtain the optimal solution on the basis of Bayesian inference, and thus it is suitable when the target function is expensive to evaluate.
For example, BO has been successful in hyperparameter tuning tasks for deep neural networks \cite{SnoekNIPS12} as each run of the training is expensive.

\emph{Preference learning} is a category of machine learning which handles preference information (\eg, $A$ is preferred to $B$) \cite{ChuICML05,Koyama:UIST:14}.
Brochu \etal\ \shortcite{BrochuNIPS07} combined BO and preference learning \cite{ChuICML05} to enable humans to perform optimization using their preferences.
We refer to this human-in-the-loop approach as \emph{preferential Bayesian optimization} (PBO)\footnote{Note that we use the term PBO in a broader sense than Gonz\'{a}les \etal\ \shortcite{GonzalezICML17}.}, and our sequential plane search is along this line.
Researchers have proposed various interaction forms of PBO \cite{BrochuSCA10,Koyama:SIGGRAPH:17,Chong:arXiv:19}.
Among them, our method is particularly inspired by the \emph{sequential-line-search} method \cite{Koyama:SIGGRAPH:17}, which was originally developed for crowdsourcing settings.
We review these previous PBO methods in detail in \autoref{sec:background}.
Also, our experiment shows that our sequential-plane-search method drastically outperforms the sequential-line-search method.
Chong \etal\ \shortcite{Chong:arXiv:19} proposed a generative image modeling method that constructs a multi-dimensional search subspace and lets the user explore it, which is similar to ours at the concept level.
Whereas their method relies on domain-specific formulations, ours is formulated as a general method so that it is applicable to various domains.
Moreover, whereas their method simply uses multiple sliders, our sequential plane search is formulated to be tightly coupled with the gallery-based interface to effectively facilitate users' design exploration.

\subsection{Optimization-Based Design}
Researchers have proposed various methods and systems for finding desired design parameters by formulating design processes as mathematical optimization problems.
In the computer graphics community, this \emph{computational design} approach has been often taken for fabrication-oriented design scenarios \cite{Umetani:SIGGRAPH:14,Prevost:SIGGRAPH:13,Bacher:SIGGRAPH:14,Li:SIGGRAPH:16,Bharaj:SA:15}.
This approach has also been taken in the human-computer interaction community for optimizing user interface design \cite{TodiDIS16,BaillyUIST13,KarrenbauerUIST14,DudleyCHI19} and building advanced creativity support tools \cite{ODonovanCHI15,Koyama:CHI:18}.
Our work is along this line, but our target problem involves handling a perceptual objective function as described in the next section.

Interactive evolutionary computation (IEC) methods \cite{TakagiProcIEEE01} also aim at involving human evaluation to produce artifacts.
In contrast to the IEC-based approach, our sequential plane search mainly aims to minimize the number of necessary queries to a human by incorporating BO techniques.
In addition, our sequential plane search is designed so that it can be performed effectively with a gallery-based interface.
Design optimization using human evaluation has also been investigated in the mechanical engineering domain \cite{Ren:JMD:11};
our approach is applicable to this domain as long as designs can be visually evaluated through a grid interface.

% !TEX root = ../paper.tex

\section{Problem Description and Background}
\label{sec:background}

In this section, we first describe our target problem using the terminologies in numerical optimization.
Then, we review the existing methods for this problem in detail.

\subsection{Problem Description}

We consider a parametric visual design task that involves $n$ design parameters.
We assume that the parameters are all continuous and thus are typically adjusted by sliders and that their visual effects are also continuous (but not necessarily linear).
The primary goal of this task is to search for the slider configuration that produces the ``best'' design (\ie, the most preferable design for the user who performs the task) among all possible designs achievable via slider manipulation.
We can interpret this task as an $n$-dimensional continuous numerical optimization problem from a mathematical viewpoint;
here, we can consider that the user's preference plays the role of the \emph{objective function} of this optimization problem and the user tries to find a maximizer of this function though exploration \cite{Koyama:OUP:18}.

Let $\calX$ be the $n$-dimensional design space that the user is going to explore (\ie, the \emph{search space}).
We assume $\calX = [0, 1]^{n}$ without loss of generality by applying normalization.
An element in this space, $\bfx \in \calX$, represents a slider configuration.
The goal of this task is to find the best parameter set, $\bfx^{*} \in \calX$.
The \emph{goodness} of a parameter set is evaluated by a perceptual function, which is called a \emph{goodness function} \cite{Koyama:UIST:14,Koyama:CHI:16,Koyama:SIGGRAPH:17}, and we represent it as $g: \calX \rightarrow \bbR$.
Then, we can write the task in the form of an optimization problem as
\begin{align}
  \bfx^{*} = \argmax_{\bfx \in \calX} g(\bfx). \label{eq:problem}
\end{align}
However, solving this problem is not easy.
The goodness function $g$ is infeasible to formulate as a simple mathematical function since the function $g$ is tightly coupled with the user's preference and even the user him/herself usually does not know what it looks like before exploring the design space $\calX$.
Thus, we cannot apply standard optimization techniques, which usually expect the objective function to be executable by computers, to this problem.
This has motivated researchers to develop human-in-the-loop optimization methods.
Note that our target problem is similar to but different from exploratory design (\eg, \cite{Talton:SA:11}), where users' preferences can change over time during the task. In contrast, we assume that the perceptual function $g$ does not change over time.

Asking a user to provide feedback about goodness requires special care;
it is not feasible for the user to reliably and consistently rate visual designs using \emph{absolute} values.
That is, we should not query the goodness value $g(\bfx)$ directly for a certain parameter set $\bfx$.
One reason is that such a value can be reliably determined only when the user is familiar with the design space $\calX$ and can imagine other possible options in $\calX$. However, this is not true in most cases, especially when the user does not have a clear goal vision at the beginning of the design process.
Also, as discussed by Brochu \etal\ \shortcite{BrochuSCA10}, \emph{drift} (\ie, the subjective scale may change over time) and \emph{anchoring} (\ie, the subjective scale may be dominated by earlier experiences) effects might cause inconsistencies.

Instead, asking about \emph{relative} preference is more promising.
The simplest form of such a preference query could be \emph{pairwise comparison}, in which the user is provided with two visual options, say $\bfx^{a}$ and $\bfx^{b}$, and then chooses his/her preferred one \cite{TsukidaTR11}.
We can interpret this information as either $g(\bfx^{a}) > g(\bfx^{b})$ or $g(\bfx^{a}) < g(\bfx^{b})$ depending on the response.
The important point is that the user can provide a preference without being familiar with the entire design space or imagining other possible options.
Also, we can expect that the drift and anchoring effects will not critically affect responses to preference queries.

Another issue that we need to consider is that human evaluation is much more expensive than typical executable objective functions.
For example, a user cannot feasibly be expected to perform a task involving $10,000$ subjective evaluations, whereas a computer can do so efficiently.
This is our motivation to use BO \cite{ShahriariProcIEEE16}, which is specifically designed to find a solution with as few queries as possible.
In the next subsection, we review previous BO-based methods for preference queries, on which our method is built.

\subsection{Preferential Bayesian Optimization Methods}

PBO is a variant of BO, which is based on preference (\ie, relative-comparison) queries instead of function-value (\ie, absolute-value) queries.
By a preference query, the PBO method obtains feedback in the following form:
\begin{align}
  \bfx^\text{chosen} \succ \{ \bfx^{(j)} \}_{j = 1}^{m},
\end{align}
where $\succ$ means that the goodness value at the left-side parameter set is likely to be larger than any goodness values at the right-side $m$ parameter sets.
We call this relational information \emph{preference data}.
The likelihood of any preference data can be modeled by the Thurstone--Mosteller model \cite{ChuICML05,BrochuNIPS07} (for $m = 1$) or Bradley--Terry--Luce model \cite{Koyama:SIGGRAPH:17,TsukidaTR11} (for any number $m$).
Note that we may omit the right-hand side curly bracket when $m = 1$ for simplicity.

Brochu \etal\ \shortcite{BrochuNIPS07} proposed using a \emph{pairwise-comparison} task (also known as \emph{two-alternative forced choice} (2AFC)), in which the user is provided with two visual options and asked to choose one.
Suppose that the method has received $i - 1$ query responses from the user and is going to determine the $i$-th query.
Let $\bfx^{+}_{i} \in \calX$ be the ``current-best'' parameter set among the observed parameter sets and $\bfx^\text{EI}_{i} \in \calX$ be the parameter set that is chosen by BO for the $i$-th iteration of the optimization.
Refer to \autoref{sec:appendix:defs} for the exact definitions of these parameter sets, but an intuition for the latter is as follows:
the point $\bfx_{i}^\text{EI}$ is defined as the point that maximizes a criterion called \emph{expected improvement} (EI), which evaluates the effectiveness of a point as the next query (we will explain this in slightly more detail in \autoref{sec:method}).
The $i$-th pairwise-comparison task is then formulated using $\bfx^{+}_{i}$ and $\bfx^\text{EI}_{i}$.
As a result of user feedback, the method obtains new preference data of either
\begin{align}
  \bfx^{+}_{i} \succ \bfx^\text{EI}_{i} \:\:\: \text{or} \:\:\: \bfx^\text{EI}_{i} \succ \bfx^{+}_{i},
\end{align}
depending on the user's choice.

Koyama \etal\ \shortcite{Koyama:SIGGRAPH:17} proposed using a \emph{single-slider-manipulation} task, in which the user is provided with a single slider and a preview widget that is dynamically updated in accordance with the slider value and asked to find the best slider tick position.
From a mathematical viewpoint, the user is considered to solve a \emph{line-search} query.
Koyama \etal\ proposed constructing the one-dimensional subspace for the $i$-th iteration, $\calS_{i}$, as
\begin{align}
  \calS_{i} = \{ (1 - t) \bfx^{+}_{i} + t \bfx^\text{EI}_{i} \mid t \in [0, 1] \}. \label{eq:line_search:subspace}
\end{align}
Then, the task for the user is described as
\begin{align}
  \bfx^\text{chosen}_{i} = \argmax_{\bfx \in \calS_{i}} g(\bfx), \label{eq:line_search}
\end{align}
and the user's feedback for this task is interpreted as
\begin{align}
  \bfx^\text{chosen} \succ \{ \bfx^{+}, \bfx^\text{EI} \}.
\end{align}
Note that more points from $\calS_{i}$ can be added into the right-side set, but this may increase the computational cost unnecessarily.
A notable advantage of this approach over the pairwise-comparison approach is that a single query can obtain important information on an additional parameter set chosen from the \emph{continuous} subspace, whereas the pairwise-comparison approach can obtain only information on \emph{discrete} parameter sets.
This makes the number of iterations necessary to obtain a good solution much smaller.

% !TEX root = ../paper.tex

\section{Approach Overview and User Interaction}

We propose a new variant of PBO called \emph{sequential plane search}, in which \emph{plane-search} queries are used for human evaluation to enable the optimization to be even more efficient.
We also propose using a \emph{zoomable grid} interface for the user to perform tasks involving plane-search queries.
We refer to the entire interactive framework consisting of the se\-quen\-tial-plane-search backend and the zoomable-grid-interface frontend as \emph{Sequential Gallery} (see \autoref{fig:teaser}).
This framework lets the user sequentially perform plane-search subtasks using the gallery-based interface to solve the target visual design optimization problem.

\subsection{Plane-Search Query}

We define a plane-search query as follows.
Let $\calP$ denote a two-dimensional manifold in the $n$-dimensional design space $\calX$ (also simply called a \emph{plane}) and $\calP_{i}$ denote the plane for the $i$-th step (we will explain how $\calP_{i}$ is parameterized and constructed in \autoref{sec:method}).
For each step, our sequential plane search asks the user to search for the best parameter set on the plane $\calP_{i}$.
From a mathematical viewpoint, the task for the user is described as
\begin{align}
  \bfx^\text{chosen}_{i} = \argmax_{\bfx \in \calP_{i}} g(\bfx). \label{eq:plane_search}
\end{align}
Note that this is analogous to a line-search query (\autoref{eq:line_search}), but our method involves two-dimensional subspaces instead of one-dimensional ones.

\subsection{Task Execution with Zoomable Grid Interface}

A straightforward way of performing the plane-search task (\autoref{eq:plane_search}) is to use two sliders mapped to the plane with a preview of the visual representation that can be dynamically updated in accordance with the slider values.
However, this approach requires the user to actively try many combinations of those two slider values at the beginning of the task to understand the design variation in the current subspace and then to adjust the slider values alternately and little by little at the fine-tuning stage.

We instead use a zoomable grid interface to execute the plane-search task.
This interface takes advantage of the fact that the subspace and display are both two-dimensional;
it displays clickable visual options in a grid so that their corresponding parameter sets are spatially mapped to the plane (see \autoref{fig:mapping}).
At the initial coarsest zoom level, the entire plane is mapped to the grid.
Once the user clicks an option, the interface goes to the next zoom level with a short zooming animation (around 1.5 seconds).
The center element of the new grid is the one chosen in the previous zoom level.
In the current implementation, we set the zooming factor as two;
that is, the new grid is mapped to an area one-quarter of that in the previous zoom level.
Note that extrapolation along the plane is performed when the user selects an option at an edge of the grid.
After a certain number of clicks (four in our implementation), this plane-search task finishes, and the method receives the parameter set that the user selected in the finest zoom level.
The procedure is illustrated in \autoref{fig:zoom}.

\figmapping
\figzoom

This interface involves only discrete selection, and thus we need to consider the effect of discretization.
The use of this interface approximates \autoref{eq:plane_search} to
\begin{align}
  \bfx^\text{chosen}_{i} \approx \argmax_{\bfx \in \calG_{i}} g(\bfx), \label{eq:discrete_plane_search}
\end{align}
where $\calG_{i} \subset \calP_{i}$ is the finite set of the parameter sets that can be accessed by the zooming procedure.
Despite the discretization of $\calP_{i}$, we can still consider that the resulting parameter set is virtually chosen from a continuous space because $\calG_{i}$ contains a large number of samples thanks to the hierarchical zooming procedure.

Thus, in a Sequential Gallery session, the user sequentially solves \autoref{eq:discrete_plane_search} for $i = 1, 2, \ldots$ until a satisfactory design is found.

% !TEX root = ../paper.tex

\section{Method: Sequential Plane Search}
\label{sec:method}

This section describes the technical aspect of the sequential plane search, which is used inside Sequential Gallery.
Note that we omit details that the sequential plane search shares with the previous methods \cite{BrochuNIPS07,Koyama:SIGGRAPH:17} and are not necessary to understand its novelty.
Those who want to implement the proposed method from scratch should refer to these papers as well.

\subsection{Plane Construction Strategy}
\label{sec:method:strategy}

To make a sequential-plane-search procedure effective, search planes must be constructed appropriately.
For the requirements for an effective algorithm, we consider two conditions as the basic design goals.
\begin{itemize}
  \item The plane should be constructed such that it is likely to minimize the number of iterations necessary for finding a satisfactory parameter set. For this, we propose a new tailored measure (\ie, a new \emph{acquisition function}) to evaluate the effectiveness of search planes (\autoref{sec:method:acquisition}).
  \item The plane should include the parameter set, $\bfx^\text{EI}$, as in the previous methods \cite{BrochuNIPS07,Koyama:SIGGRAPH:17}. This is important from a theoretical viewpoint to ensure that the plane includes the optimal solution in the ideal case that Bayesian inference is perfectly correct. Additionally, this ensures that sequential plane search always performs better than (or at least equivalent to) the previous methods because the subspaces in the previous methods are always a subset of our subspace.
\end{itemize}
We also consider two additional design goals from the user experience perspective.
\begin{itemize}
  \item The current-best parameter set, $\bfx^{+}$, should always be centered at the plane. This ensures that the position of the current-best design in the zoomable grid interaction is consistent.
  \item The plane should be \emph{planar} in the mathematical sense. That is, we do not want the plane to be curved or folded in the design space. This ensures that any set of options aligned in a direction in the grid view exhibits linear parameter changes and should help users recognize the presented subspace in many cases. See \autoref{fig:planarity} for an illustration.
\end{itemize}

\figplanarity

\subsection{Plane Construction}

Our method supposes a plane to always be a rhombus or diamond (\ie, diagonal lines are orthogonal and crossed at their centers) to avoid constructing unnecessarily skewed quadrangles.
On the basis of this, we parameterize a plane $\calP$ by its center $\bfc \in \bbR^{n}$ and two vectors $\bfu, \bfv \in \bbR^{n}$ such that the four vertices are represented as $\{ \bfc \pm \bfu, \bfc \pm \bfv \}$ (see \autoref{fig:notation}).
We represent this as $\calP(\bfc, \bfu, \bfv)$.

The basic idea to construct the plane for the $i$-th step, denoted by $\calP_{i}$, is to solve the following optimization problem:
\begin{align}
  \calP_{i} =
  & \argmax_{\calP} \hat{a}^\text{EI}(\calP; \calD_{i - 1}) \label{eq:basic:objective} \\
  & \text{s.t.}\: \left\{ \begin{array}{l} \calP \subset \calX \\ \bfx^\text{EI} \in \calP \end{array} \right., \label{eq:basic:constraint}
\end{align}
where $\hat{a}^\text{EI}$ is an acquisition function to evaluate the effectiveness of search planes, which we will define in \autoref{eq:acquisition:integral}, and $\calD_{i - 1}$ denotes the preference data obtained by the end of the $(i - 1)$-th iteration.

From a practical viewpoint, we solve the above optimization problem in a simplified manner as follows.
First, on the basis of the design goal, we always set $\bfc_i = \bfx^{+}_{i}$.
Then, we find the parameter set that maximizes the EI, $\bfx^\text{EI}$, by solving an optimization problem as in the previous methods \cite{BrochuNIPS07,Koyama:SIGGRAPH:17};
that is,
\begin{align}
  \bfx^\text{EI} = \argmax_{\bfx \in \calX} a^\text{EI}(\bfx; \calD_{i - 1}),
\end{align}
where $a^\text{EI}$ is an acquisition function to evaluate the effectiveness of search points, used in standard BO methods \cite{SnoekNIPS12};
see the appendix for details.
Note that $\hat{a}^\text{EI}$ in \autoref{eq:basic:objective} takes a search plane as an argument whereas $a^\text{EI}$ takes a search point as an argument.
Then, we set $\bfu_{i} = \bfx^\text{EI} - \bfx^{+}$, which ensures that the plane satisfies the second constraint in \autoref{eq:basic:constraint}.
The remaining one, $\bfv_{i}$, is then obtained by solving
\begin{align}
  \bfv_i =
  & \argmax_{\bfv} \hat{a}^\text{EI}(\calP(\bfc_{i}, \bfu_{i}, \bfv); \calD_{i - 1}) \label{eq:practical:objective} \\
  & \text{s.t.}\: \left\{ \begin{array}{l} \bfu_{i} \cdot \bfv = 0 \\ \bfc_{i} \pm \bfv \in \calX \end{array} \right., \label{eq:practical:constraint}
\end{align}
where the equality constraint in \autoref{eq:practical:constraint} is added for ensuring the two vectors are orthogonal and thus the plane is a rhombus.

Although the vertex $\bfx^\text{EI}_{i} = \bfx^{+}_{i} + \bfu_{i}$ is always inside the design space $\calX$, its opposite-side vertex, $\bfx^{+}_{i} - \bfu_{i}$, might be outside of $\calX$ (though the probability is very small).
If we detect this, we correct the vertex position by simply moving it along the diagonal-line direction to the closest bound of $\calX$.
Note that it is also possible not to correct the vertex position; in this case, the grid interface does not display options whose parameter sets are outside of $\calX$.

At the first step of a sequential-plane-search procedure, no preference data is available.
For this case, our current implementation simply constructs a fixed-sized square centered at the center of the search space $\calX$ with a random direction.

\fignotation

\subsection{Acquisition Function}
\label{sec:method:acquisition}

An \emph{acquisition function} evaluates the effectiveness of a search query and is used to determine the next search query so that it is as effective as possible \cite{ShahriariProcIEEE16}.
To increase the likelihood of finding a good solution, a search fquery must observe a region that is (1) likely to have a higher value (\ie, exploitation) and (2) less certain because of the lack of observations around the region (\ie, exploration).
Several different acquisition functions can be used;
among them, we choose the EI since it is commonly chosen \cite{BrochuNIPS07,SnoekNIPS12,Koyama:SIGGRAPH:17} and it balances exploitation and exploration without needing additional hyperparameter adjustment.

In the standard BO setting, in which determining a query is equivalent to finding an appropriate sampling \emph{point}, such a point is defined as a maximizer of the acquisition function \cite{ShahriariProcIEEE16}.
However, how we should determine a query in our sequential-plane-search setting is not trivial, since the query is determined by finding a \emph{plane}, not a point.
The previous sequential-line-search method \cite{Koyama:SIGGRAPH:17} finds a \emph{line} by simply connecting the current-best point and the maximizer of the acquisition function as explained in \autoref{eq:line_search:subspace}.
Although this simple approach was demonstrated to work well, it only considers the two ends of the line.

To overcome this limitation, we propose an acquisition function tailored for evaluating the effectiveness of a \emph{plane} as a search subspace for the next iteration.
We define it as a surface integral of the density of the EI value over the plane:
\begin{align}
  \hat{a}^\text{EI}(\calP; \calD) = \frac{1}{A} \int_{\calP} a^\text{EI}(\bfx; \calD) dS, \label{eq:acquisition:integral}
\end{align}
where $A$ is the area of the plane $\calP$.
In practice, we approximate it by a summation at $N$ sampling points on the plane, $\{ \bfx^{(j)} \}_{j = 1}^{N}$.
More specifically,
\begin{align}
  \hat{a}^\text{EI}(\calP; \calD) \approx \frac{1}{N} \sum_{j = 1}^{N} a^\text{EI}(\bfx^{(j)}; \calD). \label{eq:acquisition:approximation}
\end{align}
In the current implementation, we use $N = 25$ points simply sampled in a 5-by-5 lattice pattern as shown in \autoref{fig:notation} (Right).

\subsection{Interpreting Query Response as Preference Data}

Given a plane $\calP_{i}$, the user provides the maximizer on the plane, $\bfx^\text{chosen}_{i}$, as feedback to our method.
Then, our method interprets this information as the following preference data:
\begin{align}
  \bfx^\text{chosen}_{i} \succ \{ \bfc_i, \bfc_i \pm \bfu_i, \bfc_i \pm \bfv_i \}.
\end{align}
Note that more parameter sets from $\calP_{i}$ can be added to the right-side sets.
However, to avoid an unnecessarily large computational cost, we use the above five parameter sets as the representatives of the plane.

\subsection{Implementation Details}

\figphotofull

We implemented BO based on a \emph{Gaussian process} \cite{Rasmussen05} prior with the \emph{automatic relevance determination (ARD) Mat\'ern 5/2} kernel, following the suggestion by Snoek \etal\ \cite{SnoekNIPS12}.
We determine the kernel hyperparameters in the same way as the previous paper \cite{Koyama:SIGGRAPH:17}:
we use \emph{maximum a posteriori} (MAP) estimation every time new preference data is added and assume a log-normal distribution $\calL\calN(\mu, \sigma^{2})$ as the prior distribution for each kernel hyperparameter.
We set $\mu = 0.2$ for the \emph{amplitude} and $\mu = 0.5$ for every \emph{length scale}, and $\sigma^{2} = 0.01$ for both;
refer to Snoek \etal\ \shortcite{SnoekNIPS12} for the definitions of these hyperparameters.
We handle the equality constraint in \autoref{eq:practical:constraint} by simply interpreting it as a soft constraint term:
\begin{align}
  - (\bfu_{i} \cdot \bfv)^2,
\end{align}
and adding it to the objective function.
As a result, the problem can be simply considered as an unconstrained, bounded optimization problem, and the overall objective function can still be differentiable with respect to the unknown, $\bfv$.
Thus, we solve it by using the limited-memory BFGS (L-BFGS) method \cite{Nocedal06}.
Note that, to handle the equality constraint more exactly, the \emph{augmented Lagrangian} method \cite{Nocedal06} can be used here.
However, we decided to use the simplified approach as we found it sufficiently accurate and also efficient.
As the problem of Equations \ref{eq:practical:objective} and \ref{eq:practical:constraint} can have multiple local maxima, we perform the optimization 10 times with random initial solutions in parallel and then use the best-found solution.

% !TEX root = ../paper.tex

\section{Applications}

An advantage of our Sequential Gallery is that it does not rely on any domain-specific formulation and thus can be immediately applied to many visual design domains involving a set of continuous parameters.
To demonstrate its generality, we created a photo color enhancement system and a human body shape generation system.

\subsection{Photo Color Enhancement (12 Design Parameters)}
\label{sec:app:photo}

Photo color enhancement involves editing colors in a photograph to make the photograph more appealing.
For this task, in addition to basic parameters (brightness, contrast, and saturation), we implemented an advanced color balance functionality that independently manipulates colors in shadows, midtones, and highlights.
Since this advanced color balance is important, popular photo enhancement tools \cite{InstagramEffects,AdobePhotoshopColorBalance} usually have it or similar functionality.
In total, this system has 12 design parameters to be adjusted.
Note that the expressiveness of our implementation is much higher than that of the previous work \cite{Koyama:SIGGRAPH:17}, in which only basic $6$ parameters were implemented.

\autoref{fig:photo-full} (Left) shows a full sequence of a Sequential Gallery session for photo color enhancement.
In this example, the user obtained a satisfactory result after $4$ iterations.
\autoref{fig:photo-full} (Right) shows additional results, and refer to the supplemental video figure for corresponding sequences and a larger view.

\subsection{Human Body Shape Design (10 Design Parameters)}

It is useful for end-users to be able to easily design human body shapes for various purposes such as those for creating virtual avatars and for visually conveying body shapes in their mind to someone else.
To enable this, we used the skinned multi-person linear (SMPL) model \cite{LoperSA15}, which is a publicly available pre-trained human shape model based on principal component analysis.
We used its first 10-dimensional latent parameters and set the bound of the design space as three times the standard deviation from the mean shape.
\autoref{fig:smpl-space} shows variations of human body shapes achievable by this system.

Compared with photo color enhancement, human body shaping requires comparing more details of visual representations among options, and we noticed that cells in a 5-by-5 grid with a 13-inch display were too small for this application.
Thus, we used a 3-by-3 grid for the interface of this application (see \autoref{fig:body-results} (Left)).

One challenge in this design domain is that the latent space of the learned model does not have \emph{semantically meaningful} dimensions, so exploration by direct slider manipulation is difficult.
For this, \emph{Body Talk} \cite{StreuberSIGGRAPH16} remapped these parameters to a semantic space by crowdsourcing perceptual annotations.
Our method takes a different approach:
we handle the design space as a black-box and use the WYSIWYG interface.

An advantage of involving users in the loop is that users can produce designs even from vague pictures in their minds.
Following the previous work \cite{StreuberSIGGRAPH16}, we created a human body shape from a description in a novel, \emph{The Maltese Falcon}:
\begin{quote}
  \textit{He was of medium height, solidly built, wide in the shoulders, thick in the neck, with a jovial heavy-jawed red face $\ldots$} \cite{Falcon}
\end{quote}
\autoref{fig:body-results} (Center) shows the result obtained after $7$ iterations.
We also created a body shape of a famous fictional character, \emph{Spider-Man} \cite{SpiderMan}, without looking at any references.
\autoref{fig:body-results} (Right) shows the result obtained after $10$ iterations.
Another interesting usage is to reproduce body shapes in photographs \cite{StreuberSIGGRAPH16}.

\figsmplspace
\figbodyresults

% !TEX root = ../paper.tex

\section{Evaluation}

\subsection{Experiment Using Synthetic Functions}

First, we conducted an experiment using synthetic functions to artificially simulate users' responses on the basis of their preferences.
This simulation-based approach is useful to generate a large number of responses to properly understand the behavior of our sequential plane search from a technical viewpoint.

\subsubsection{Goals}

The specific goals of this experiment were
(1) to evaluate the efficiency of the proposed sequential plane search and
(2) to validate whether the use of the BO-based plane construction is effective.

\subsubsection{Methods to be Compared}

We compared three methods:
\begin{itemize}
  \item \textbf{SLS}: The sequential-line-search method \cite{Koyama:SIGGRAPH:17} that constructs a one-dimensional subspace using BO. Except for the subspace construction, SLS used the same settings as the implementation of our method, such as kernel function choice and hyperparameter handling.
  \item \textbf{SPS (Random)}: The sequential-plane-search method as proposed in this paper, but using a random plane construction. More specifically, a plane is constructed by first setting $\bfc = \bfx^{+}$ like in our proposed method, and then choosing $\bfu$ and $\bfv$ randomly such that they have a constant length ($\| \bfu \| = \| \bfv \| = 1$) and are orthogonal ($\bfu \cdot \bfv = 0$).
  \item \textbf{SPS (Ours)}: The sequential-plane-search method using the proposed BO-based plane construction.
\end{itemize}
Note that we omitted the BO-based pairwise-comparison method proposed by Brochu \etal\ \shortcite{BrochuNIPS07} because Koyama \etal\ \shortcite{Koyama:SIGGRAPH:17} already reported that it consistently requires much more iterations than SLS.

\subsubsection{Synthetic Functions}

We used two different synthetic functions.
The first consists of a single isotropic (\ie, no covariance) Gaussian kernel:
\begin{align}
  g^\text{synth}_{1}(\bfx) = \exp \left[ - (\bfx^{*} - \bfx)^\T (\bfx^{*} - \bfx) \right], \nonumber
\end{align}
where we set $\bfx^{*} = \bmat{ 0.3 & \cdots & 0.3 }^\T$.
The second is a test function called the \emph{Rosenbrock function} \cite{SciPyRosenbrock}.
We take its negative to obtain maximizers rather than minimizers since our formulation (\autoref{eq:problem}) uses maximization.
Also, we scale it by a factor of $0.25$ so that its important region lies within $\calX = [0, 1]^{n}$.
As a result, it is defined as
\begin{align}
  g^\text{synth}_{2}(\bfx)
  = - \sum_{i = 1}^{n - 1} \left[ 100 ( 4 x_{i + 1} - 16 x_{i}^{2} )^{2} + \left( 1 - 4 x_{i} \right)^{2} \right], \nonumber
\end{align}
which has its global minimum at $\bmat{ 0.25 & \cdots & 0.25 }^\T$.
The first function is relatively easy to optimize whereas the second is much more difficult because of its complex shape.
We measured the performance by using the \emph{optimality gap}:
the difference between the optimal function value and the best-found function value \cite{WangJAIR16}.

\subsubsection{Results}

\autoref{fig:synthetic} shows the results of 50 trials for each method and for each function.
We can observe that the SPS methods are consistently far superior to the SLS method in all the settings.
This indicates that using plane-search queries instead of line-search queries significantly reduces the number of iterations necessary for reaching good solutions.
Also, we can observe that our SPS is consistently superior to the random SPS except for the first few iterations.
This indicates the importance of taking the BO-based plane construction approach;
it clearly improves the performance after several iterations.

\figsynthetic

To determine whether these differences in performance are statistically significant, we performed two-sided Mann--Whitney \textit{U} tests ($\alpha = 0.05$) to compare the three methods.
As this test involves multiple comparisons, we adjusted the significance level $\alpha$ in each comparison by the Bonferroni correction.
The performance of the SLS method was significantly different from those of the SPS methods in all the functions at all the iteration counts.
The performances of the two SPS methods were not significantly different for the first 1, 6, 4, and 10 iterations (for Isotropic Gaussian 5D, Isotropic Gaussian 15D, Rosenbrock 10D, and Rosenbrock 20D, respectively) but became significantly different at the next iteration counts ($p \ll 0.001 \: (f = 0.803)$, $p \ll 0.001 \: (f = 0.763)$, $p = 0.001 \: (f = 0.688)$, and $p = 0.001 \: (f = 0.694)$, respectively, where $f$ represents the common language effect size);
the reason for the lack of significant differences in the first few iterations could be that the observed data points were so sparse given the high dimensionalities that the BO-based SPS performed almost pure exploration like the random method.
After those iteration counts, the differences continued to be statistically significant, but there was one exception that the differences were not significant in the Isotropic Gaussian 5D function after 15 iterations;
the reason of this exception could be that both methods had sufficiently approached the optimal solution and thus already converged.

\subsection{Preliminary User Study}

The first evaluation validated the effectiveness of our sequential plane search in goal-driven settings by using synthetic functions.
However, the effectiveness of the overall interactive framework remains unevaluated.
Thus, we conducted an additional small user study, where we used the photo color enhancement scenario described in \autoref{sec:app:photo}.

\subsubsection{Goals}

The specific goals of this user study were (1) to evaluate whether even novice users who are not necessarily familiar with the target parameters and not likely to have clear mental images at the beginning of the task can perform the plane-search subtasks by the zoomable grid interface and find a satisfactory parameter set, (2) to evaluate whether the interface can facilitate exploration and provide inspiration, and (3) to gather qualitative feedback on our interactive framework.

\subsubsection{Method}

Five students and one researcher ($P_{1}, \ldots, P_{6}$) participated in this user study.
This study consisted of four parts: an instruction session, the first photo enhancement task, the second photo enhancement task, and a questionnaire.
We prepared three photographs and used one of them for instruction for every participant and the other two for the main tasks.
For photo enhancement, we instructed the participants to imagine that they were going to upload the photographs to their Facebook or Instagram accounts and wanted to make the photographs appealing to their friends.
We asked them to continue each task for $15$ iterations (\ie, a sequence of $15$ plane-search subtasks) regardless of whether they were already satisfied with intermediate results or not.
For recording purposes, we also asked the participants to push a ``satisfied'' button on the screen during the iterations of each task when they felt satisfied with the current design (\ie, the option clicked last).
We used a 13-inch MacBook Pro with a mouse and maximized the application window size.
After the tasks, we asked the participants to fill in a questionnaire consisting of a question about expertise in photo color enhancement, questions arranged on a $7$-pt Likert scale with $7$ corresponding to ``strongly agree,'' and an optional free comment space.

\subsubsection{Results}

Five participants (other than $P_{3}$) pushed the button to indicate that they were satisfied within $15$ iterations in the main tasks.
The participant ($P_{3}$) who did not push the button in one of the main tasks informally told us that the initial photograph was already satisfactory and it was not clear whether the button should be pushed at the beginning.
Overall, these results sugget that the framework could provide satisfactory results.
The mean iteration count necessary for finding satisfactory results was $5.36$ with $SD = 2.69$ (the task in which the button was not pushed was excluded).
One plane-search subtask took $14.8$ seconds on average.

In the questionnaire, $P_{4}$ described him/herself as an expert and the other five described themselves as novices.
For the statement, \qt{I could find satisfactory photo enhancement results,} the mean score was $5.83$ with $SD = 0.753$.
For another statement, \qt{I could get inspiration for possible enhancement from the grid view,} the mean score was $6.50$ with $SD = 0.548$.
Overall, these results support our claims.

We obtained feedback that validates our interface design.
$P_{2}$ wrote that he/she selected designs \qt{based on criteria that I didn't have at the beginning} thanks to the encouragement of our framework to explore designs.
Also, $P_{3}$ wrote the interface \qt{is really inspiring in that it makes me want to try out different styles,} which validates that our interface can facilitate exploration.
$P_{4}$ appreciated that \qt{the system proposed some nice photos} and so that he/she \qt{could get inspiration.}
We also obtained feedback for improvement.
$P_{4}$, who self-described as an expert, wrote that he/she was familiar with photo enhancement parameters and so they wanted a direct parameter manipulation functionality along with the gallery-based search functionality.
$P_{3}$ wanted to \qt{have the original photo alongside to compare with during the enhancement.}

% !TEX root = ../paper.tex

\section{Discussions and Future Work}

\paragraph{Initial Plane Selection}
Our current implementation randomly chooses the initial plane since no preference data is available at the beginning.
Another possible initialization method is to insert a step similar to Design Gallery \cite{MarksSIGGRAPH97} before running the sequential-plane-search procedure as follows.
First, the design system provides a diverse set of options from the entire design space by embedding them into a two-dimensional widget and then let users choose the best one.
Then, the system handles this preference data (\ie, the chosen option is preferred over the other options), $\calD_{0}$, by the Bradley--Terry--Luce model \cite{TsukidaTR11} and then calculates the acquisition function (\autoref{eq:line_search:subspace}) to construct the initial plane.
Finally, the system begins sequential plane search with the initial plane and also with the additional preference data $\calD_{0}$.
Note that the Design Gallery approach is not for exploring a multi-dimensional space to sequentially refine a solution, though it is good at providing an overview of the entire design space and letting users quickly pick out an initial solution.
Our Sequential Gallery can thus complement the Design Gallery approach.

\paragraph{Further Understanding of Interfaces}
We adopted the grid interface to help users not only find the best parameter set within the subspace but also easily grasp the landscape of the current subspace and obtain inspiration from the interface.
Thus, the goal was not to make each task execution quicker.
Nonetheless, it would be interesting to perform comparative studies to investigate both qualitative and quantitative differences between interfaces (a single slider, two sliders, and ours).
Also, as our user study was only preliminary, formal studies will need to be conducted with more participants and more practical settings.

\paragraph{Grid Resolution and Zooming Levels}
We used fixed values for the grid resolution (\eg, 5-by-5 for the photo enhancement) and the number of zooming levels for each query (\eg, four).
We empirically set these values for each application since every design domain has different appropriate values for these variables.
Nonetheless, automatically determining appropriate values is important future work.
For example, it is worth investigating how to dynamically adjust the number of zooming levels by analyzing the \emph{just-noticeable difference} (JND) of the visual representations in the grid.
Another direction is to enable users to interactively adjust these values during tasks to avoid unnecessarily fine zooming levels or grid resolutions.

\paragraph{Plane Construction Strategy}
Our strategy always chooses the maximizer of the acquisition function, $\bfx^\text{EI}$, as one of the vertices of the rhombus.
While we believe this strategy is reasonable in the sense that we discussed in \autoref{sec:method:strategy}, it is worth investigating other strategies that provide better exploration-exploitation balancing or theoretical regret bounds \cite{Srinivas:InfTheory:12}.
Also, our strategy enforces the current-best parameter set, $\bfx^{+}$, to be the center of the next search plane, which is for ensuring consistent interaction when moving from one search plane to another.
Another possibility for improving usability would be to introduce a constraint for enhancing continuity between search planes so that users can more easily understand variations in new search planes.

\paragraph{Non-Visual Design}
One notable limitation of Sequential Gallery is that it may be ineffective with non-visual designs such as electronic timbre design for a sound synthesizer.
This limitation is because the grid interface assumes that designs are visually recognizable at a glance.
Another limitation is that our sequential plane search does not handle discrete parameters such as layouts, fonts, or filter types.

\paragraph{Even Higher Dimensionality}
BO is known to perform poorly with very high dimensionality (\eg, over 20 dimensions) \cite{WangJAIR16}.
However, many design tasks involve such high-dimensional design spaces.
For example, facial expression modeling for a virtual character usually involves around 50 parameters and sometimes around 1,000 \cite{LewisEG14};
to conduct these tasks with our sequential plane search, users need to choose a moderate number of relevant parameters beforehand.

\paragraph{Prior Knowledge}
Our method assumes that everywhere in the search space is equally good (or bad) at the beginning of the process.
To accelerate the search, incorporating prior knowledge about the target design domain would be beneficial.
For example, we could build a rough approximation of the goodness function by gathering preference data by crowdsourcing \cite{Koyama:UIST:14} or implementing common practices in the domain and then use it as a prior of the Bayesian inference.

\paragraph{Time-Varying Preference}
Our method handles all preference data equally to infer latent preferences and determine search planes.
This assumes that users' preferences do not change over time.
In practice, however, this assumption is not always valid (\cf, exploratory design).
We could accept such concept drift by simply allowing users to discard accumulated preference data (either entirely or partially) at any time during the search.
Incorporating the time-varying property into the BO formulation is also interesting future work.

\paragraph{Acquisition Function Choice}
Following previous work, we chose the EI as the criterion to evaluate the effectiveness of a search point and then proposed its extension for evaluating the effectiveness of a search plane (\autoref{eq:acquisition:integral}).
Other acquisition functions (\eg, Gaussian process upper confidence bound \cite{Srinivas:InfTheory:12}) are applicable for plane search in the same way.
Note that there exist acquisition functions specifically tailored for the \emph{discrete} pairwise-comparison setting \cite{GonzalezICML17}.
However, they are not directly applicable to our problem as our goal is to evaluate the effectiveness of a \emph{continuous} subspace.
This is why we proposed an integral-based acquisition function.

\paragraph{Sequential Subspace Search}
Line search (\autoref{eq:line_search}) and plane search (\autoref{eq:plane_search}) are one- and two-dimensional, respectively.
This suggests a generalization:
letting users perform $m$-dimensional search subtasks sequentially to solve the original $n$-dimensional problem ($m < n$), which we call \emph{sequential subspace search}.
One of our findings is that the convergence with $m = 2$ is drastically faster than that with $m = 1$ (\autoref{fig:synthetic}) while plane-search tasks remain easy thanks to the zoomable grid interface.
We expect that the convergence will become even faster with $m \ge 3$.
However, the subtasks would become unreasonably tedious since active trials and errors are inevitable for users to understand $m$-dimensional subspaces.
This is why we chose $m = 2$ and investigated an interface suitable for this choice.

\paragraph{Latent Spaces of Deep Generative Models}
Recent advances in deep generative models have demonstrated a new paradigm of design, in which users obtain various designs by specifying parameter sets in latent spaces.
However, these latent spaces are difficult for users to explore because they are black-boxes for users and usually intractably high dimensional (\eg, 128 dimensions for geometry generation \cite{ChenCVPR19}), which poses a new problem that needs to be solved by computational techniques and interaction designs in combination.
We believe that this work could be an important step in this direction.

\section{Conclusion}

We presented sequential plane search, a novel optimization method for parametric visual design tasks.
This method involves the user in the loop of its procedure;
it decomposes the target high\--di\-men\-sion\-al problem into a sequence of much easier two-dimensional subtasks, which can be solved by the user via a simple zoomable grid interface.
Our experiment using synthetic functions revealed that using plane-search subtasks was much more effective than using line-search subtasks \cite{Koyama:SIGGRAPH:17} and that our BO-based plane construction was significantly more effective than a random plane construction.
In addition, our user study confirmed that novices could perform our sequential plane search via the zoomable grid interface and find satisfactory results in the photo color enhancement scenario.
The study also confirmed that the interface could facilitate exploratory design.

The overall framework, called Sequential Gallery, is quite general and does not rely on any domain-specific formulations, which makes it directly applicable to various problems.
Furthermore, it provides a promising future opportunity to adapt our framework to specific problems by incorporating domain-specific considerations into the BO routine (\eg, \cite{Chong:arXiv:19}) or the interface design.
We plan to make our source codes accessible to encourage this direction.

\begin{acks}
This work was supported in part by \grantsponsor{JST}{JST}{} ACCEL Grant Number \grantnum{JST}{JPMJAC1602}, Japan, and \grantsponsor{JST}{JST}{} CREST Grant Number \grantnum{JST}{JPMJCR17A1}, Japan.
\end{acks}

\bibliographystyle{ACM-Reference-Format}
\bibliography{references,my-papers}

\appendix

% !TEX root = ../paper.tex

\section{Definitions of $\bfx^{+}$ and $\bfx^\text{EI}$}
\label{sec:appendix:defs}

Let $\mu : \calX \to \bbR$ and $\sigma^{2} : \calX \to \bbR_{\geq 0}$ be functions that return the mean and variance, respectively, of the posterior distribution of the goodness value inferred by Gaussian process regression using the currently available data.
We denote by $\bfx^{+} \in \calX$ the ``current-best'' parameter set among the observed parameter sets.
A possible criterion for the ``best'' here is the $\mu$ value \cite{BrochuNIPS07,Koyama:SIGGRAPH:17} and we followed this approach in the evaluation.
Another possibility is to simply use the user's last selection as $\bfx^{+}$, which is actually more stable sometimes.
We denote by $\bfx^\text{EI} \in \calX$ the parameter set that maximizes the EI value, which is calculated by
\begin{align}
  a^\text{EI}(\bfx) = \sigma(\bfx) (\gamma(\bfx) \Phi(\gamma(\bfx))) + \phi(\gamma(\bfx))) \text{ if $\sigma(\bfx) > 0$ else $0$,}
\end{align}
where $\gamma(\bfx) = (\mu(\bfx^{+}) - \mu(\bfx)) / \sigma(\bfx)$, $\phi$ is the standard normal function, and $\Phi$ is the cumulative distribution function of $\phi$ \cite{SnoekNIPS12}.

\end{document}